\documentstyle[11pt]{article}
\title{\vspace{2cm}\Large\bf \begin{center}
Why is the Hubble flow so quiet?
\end{center}
\begin{center}
 \author{
  Arthur Chernin(1,2), Pekka Teerikorpi(2), and Yurij Baryshev(3) \\
 (1) Sternberg Astronomical Institute,\\
 Moscow University, 119899, Moscow, Russia,\\
 (2) Tuorla Observatory, \\
 University of Turku, FIN-21500 Piikki\"o, Finland,\\
 (3) Institute of Astronomy, St.Petersburg State University,\\
 Staryj Peterhoff,198504, St.Petersburg, Russia.\\
 }
\end{center}}

\date{~}

\begin{document}

\maketitle

\begin{abstract}
The cosmological vacuum, which is perfectly uniform, dominates by
density  over all the forms of cosmic matter. It makes the
Universe be actually more uniform than it could be seen from the
visible picture of the highly non-uniform matter distribution,
especially inside the observed cell of uniformity ($100-150$
Mpc). This uniformity reveals itself in the structure of the Hubble
matter flow which extends over a giant range of cosmic space
scales -- from few Mpc to a thousand Mpc, -- preserving its
kinematical identity. According to Sandage (1999), this flow is
mysteriously   regular and quiet even deep inside the cell of
uniformity. An answer we propose to the question in the title
above is as follows: This is most probably because the flow is
dynamically controlled by the cosmological vacuum. An additional
conjecture of cosmological intermittency, that addresses a complex
statistical structure of initial chaotic perturbations, is also
suggested in this context.
\end{abstract}

\section{Introduction}

A new development in cosmological studies has been stimulated by
the recent discovery of the cosmic anti-gravitating vacuum and the
measurements of the vacuum energy density derived from supernova
Type Ia observations (Riess et al. 1998; Perlmuter et al. 1999).
It is surprising that the vacuum density proves to exceed the
observed mean matter density:
$\Omega_ {vac} = 0.7 \pm 0.1$, while $\Omega_{mat} = 0.3 \pm 0.1$.
It is even more puzzling that the
two densities are so close to each other, on the order of
magnitude, at this particular stage of the cosmological
evolution; perhaps this poses the most serious challenge to
current cosmological concepts.

On the other hand, the discovery of the vacuum seems to offer new
promising approaches to other important theoretical problems,
both classic and modern ones. We discuss here only one aspect of
the new situation in cosmology, which concerns the major topic of
this meeting, -- this is the uniformity of the observed Universe.

The fact that the vacuum is perfectly uniform and dominates by
density over all the forms of cosmic matter means that the
Universe is actually more uniform than what the cosmic matter
distribution indicates. Indeed, because of the vacuum, the
total (matter + vacuum) density of the Universe appears smoother,
and, as a result, the space geometry of the Universe proves to be
closer to that of the Friedmann model, at the present stage of
the cosmological expansion. We expose below inferences, coming
from this fact, with an emphasis on the observed kinematical
structure of the Hubble expansion flow.

\section{The Hubble enigma}

As is well-known, in 1917, when Einstein proposed his first cosmological
model, which was a model of a static uniform Universe, the cosmological
expansion was already discovered. This was done by Slipher who published
his report in the same year 1917. Seven years after, in 1924,
Friedmann discussed Slipher's observations and treated them as an
evidence for his model of non-static, expanding Universe.
Friedman model predicts the linear velocity-distance relation in
the expanding uniform Universe. This relation was found by Hubble,
and the results of his observations were published in 1929.

 In his five-page paper of 1929, Hubble reported
the discovery of `a relation between distance and radial velocity
among extra-galactic nebulae' which is the now-famous Hubble law
of the cosmological expansion. If to look at the V vs. R plot
given in the Hubble (1929) paper, one may see that the maximal
velocity observed is slightly above 1000 km/s. With the commonly
accepted now Hubble constant $H = 50-65$ km/s/Mpc, the
corresponding distances are not larger than 20 Mpc.

Hubble himself believed in 1929, as is well-known, that the
distance limit in his observations is about 2 Mpc -- see again
the same plot. But it is more significant that, these distances,
be they 2 or 20 Mpc, are not the distances of cosmological scale.
Standard text-books tell us that cosmology begins at distances
that are larger than the size of the cell of uniformity which is
$ 100-150$ Mpc or more. According to the standard cosmological
arguments, the dynamics described by the Hubble law, $V = H\times
R$, is a consequence of a uniform distribution of
self-gravitating matter, and the standard model leading to the
linear velocity field is only valid for $R$ which are larger than
the size of the cell of uniformity.

If to look again at the same V vs. R plot, one may realize that
the cosmological expansion with the Hubble linear law starts very
near us: the smallest distances in the plot are just few Mps, with
re-scaling to the present-day value of the Hubble constant.

After all this, one may ask: Does the `relation between
distance and radial velocity among extra-galactic nebulae'
observed within 2-20 Mpc have any cosmological relevance?

The seemingly incompatible co-existence of the regular Hubble
flow with the highly non-uniform matter distribution deep inside
the cell of matter uniformity is what we call the Hubble enigma.

\section{Seventy years later}

Then, let us address another paper on the subject that was
published in 1999, 70 years after the Hubble paper. Sandage
(1999) analyzes the rate of cosmological expansion which is
measured by the Hubble constant (see also Sandage, 1986, Sandage
et al. 1972) using velocity measurements up to hundreds of thousand
km/s and the most reliable distance determinations. The global rate
which is appropriate to the space scales well beyond the cell of
uniformity  ($100-150$ Mpc) is studied as well as the very local
velocity field and the local rate of expansion. `Very local'
means very near the Local Group, down to just $R = 1.5-2$ Mpc.

The rate of expansion is determined in the Virgocentric frame and
obtained with the use of special corrections which are  not only
the correction for the Sun motion used by Hubble, but also the
correction for the galaxy motion in the Local Group and the Virgo
infall. Sandage's `strong conclusion is that the local rate is
similar, if not precisely identical, to the global rate at the
$\sim 10$\% level'. (See for comparison  Karachentsev and Makarov,
1996, 2000, Wu et al., 1999, Zehavi, Dekel, 1999, Ekholm et al.,
1999).

To summarize the present-day observational situation after Sandage
(1999),

1) the fact of expansion in the local volume with the sizes from
1.5-2 to 20 Mpc discovered by Hubble has again been  confirmed;

2) the kinematical identity of the local ($< 20$ Mpc) expansion
with the global ($>> 100-150$ Mpc) expansion has been proved with
a high accuracy.

It is also important (and seems even more puzzling) that -- after all the
corrections -- the recent observations of the local velocity field, based
on cepheid-distances, have indicated
that

3) the linear Hubble flow is very quiet in the local volume: the
local velocity dispersion about the mean local expansion rate is about or
smaller than 60 km/s.

As Sandage (1999) says, "explanation of why the local expansion
field is so noiseless remains a mystery", and again in the same
paper: `We are still left with the mystery...'

\section{Vacuum and local volume}

To start a theoretical discussion, note first that one-to-one
relation between matter motion and density distribution is
possible in one and only one case, namely when the matter
dynamics is self-consistent, which means that matter itself (both
visible and dark matter) is the only dynamical factor that
controls its motion. Since such a one-to-one relation is not
really observed, one arrives at a conclusion like that:  there is
seemingly an additional dynamical factor that affects the matter
motion on the space scales from few Mpc to 20 Mpc and further to
the largest observed distances. This cosmological factor, which is
external to matter itself, makes the matter motion be more
regular than it might be expected from the data on matter density
distribution. And this is because of this dynamical factor that
the highly non-uniform matter distribution in the local volume
can have fairly regular motion in this volume. This factor must be
stronger dynamically than the self-gravity of matter on the
distances of few Mpc and larger.

The only reasonable candidate to this role is obviously the
cosmological vacuum.

Historically, after Einstein introduced the cosmological constant
$\Lambda$, Lema\^{i}tre was the first to associate $\Lambda$ with
the vacuum having positive energy density ($\rho_{vac}c^2$) and
negative pressure ($-\rho_{vac}c^2$) leading to a repulsive
cosmological force. In the 60s, Gliner made it clear that the
medium with the equation of state $\epsilon = - p$ described by
the cosmological constant has the major mechanical property of
vacuum: rest and motion can not be discriminated relative to this
medium. In other words, any matter motion is co-moving to this
medium.

With the vacuum density derived from distant supernovae
observations, $\rho_{vac} \simeq 0.7 \rho_{crit}
 = 4.7 \times 10^{-30}h_{60}^2$ g/cm$^3$,
we may try to reconsider the
matter dynamics in the local volume as controlled, in general, by
baryonic luminous matter, dark
matter and the cosmological vacuum. For the scales larger than
$\simeq 0.5$ Mpc, the mean density of dark matter is over ten times
greater than the mean density of baryons (see for a review
Bahcall et al., 1995),  so on these
 scales the dynamics is actually determined by the
competition of the gravity of dark matter and the anti-gravity of
the vacuum. On such scales, the mass-to-luminosity ratio seems to
be more or less constant, hence the dark matter distribution may
be traced more or less closely by the luminous matter. As the
mean density of luminous matter decreases with increasing scale,
so that $M \propto R^D$  (where $D = 1-2$), also dark matter may
follow the same law (see  Peebles, 1993, Teerikorpi, 1997,
Teerikorpi et al., 1998, Baryshev et al., 1998).

If so, on which scales the vacuum density starts to dominate
over dark matter density?

For a simple estimate of the characteristic scale beyond which
vacuum dominates, we may consider two models that could imitate
the real picture in the local volume.  In Model A, a spherically
symmetric distribution of matter with mass $M(R)$ is placed on
the background of the vacuum. And in Model B, two spherical
matter halos of equal mass $M(R)$ are on the same background.
Model A represents a smooth version of dark matter distribution,
while Model B deals with a maximally clumpy distribution.

These models are described by the dynamical equation:

\begin{equation}
d^{2} R/d t^2 = - G M_{eff}/R^2; \;\;\; M_{eff} = M(r) - (4\pi/3)
2 \rho_{vac} R^3.
\end{equation}
where $M(r) = M_* (R/R_*){D}$, $R(t, M)$ is the radius of mass
distribution in Model A and the distance between the centers of
the halos in Model B. The vacuum density $\rho_{vac}$ relates to
Einstein's cosmological constant $\Lambda$ as $\Lambda = 8\pi
G\rho_{vac}/c^2$.

The vacuum term in this equation dominates over the matter term at distances
\begin{equation}
R > R_{vac} = R_*^{-D/(3-D)} M_*^{1/(3-D)} (3/8\pi \rho_{vac})^{1/(3-D)}.
\end{equation}

The appropriate mass here is $M_* =2\times 10^{12} M_{\odot} $
which is the most recent estimate for the mass of the Local
Group, and the radius $R_* = 1$ Mpc may be considered as the
corresponding size of the matter distribution (van den Bergh
19990). Then the critical distance $R_{vac}$, where the vacuum
starts to dominate,
 is  $R_{vac} \simeq 2$ Mpc for $D = 1$ and $R_{vac} \simeq 4$ Mpc
for $D = 2$. Hence, the models put $R_{vac}$ robustly
into the range between 2 and 4 Mpc.

This figures prove to be close to the distance $R = 1.5-2$ Mpc,
where the Hubble law emerges, according to Sandage (1986, 1999) and
Ekholm et al. (1999).

Is it hardly a pure coincidence. This is rather an
indication that the kinematics of the local matter flow may indeed be
controlled mainly by the uniform vacuum.

If so,
the rate of expansion -- in the zero (and main!) approximation -- is
expressed via the vacuum density and therefore
the rate is the same at all distances starting at 2 -- 4 Mpc:
\begin{equation}
H = (dR/dt)/R = \sqrt{(8 \pi G/3) \rho_{vac}}.
\end{equation}

Generally, the Universe is uniform -- in the zero approximation
-- on such distances from us, where the uniform vacuum
dominates by effective density over the non-uniform matter; on
these distances, the total cosmological density $\rho =
\rho_{vac} + \rho_{dark-matter} + \rho_{baryomic}$ may be
considered as uniform one, in this approximation. In the
cosmological environment of the Local Group, this zero-order
uniformity, as we see, begins with the distances of 3-6 Mpc, so
that we may say now that it is from these scales that cosmology
begins. Perhaps this way the Hubble enigma may find its solution,
at least in principle.

Note that the critical distance $R_{vac}$ we find
above does not deviate much from the minimal distance given by both
Hubble and Sandage for the Hubble law. The discrepancy will
be smaller, if one uses a larger value for the vacuum density, --  say,
$\Omega_ {vac} = 0.8$ and  $\Omega_{mat} = 0.2$, or even
$\Omega_ {vac} = 0..9$ and  $\Omega_{mat} = 0.2$; in the
latter case, $\Omega = \Omega_ {vac} + \Omega_{mat} > 1$. In this sense,
may the observed minimal distance $R = 1.5- 2$ Mpc indicate that the
vacuum density is indeed larger than one accepted today?

In other cosmological environments, the dominance of the vacuum my
start from different, smaller or larger, distances. For instance,
around the Coma cluster, the vacuum dominates starting from
$20-30$ Mpc or so. For a cosmologist located within the Coma
cluster, cosmology begins with the distances of 30 Mpc.

\section{Vacuum and stability of the flow}

According to Sandage (1999), the local expansion flow is
surprisingly cold: the characteristic velocity dispersion, $\sigma
\le 60$ km/s, in the local volume (Sec.3). From the theoretical
point of view, the quietness of the expansion flow controlled
mainly by the vacuum may be understandable, -- at least in
principle. Indeed, non-uniform matter distribution has only a weak
effect on the matter flow when the self-gravity of matter is
weaker than the anti-gravity of the uniform vacuum. The
vacuum-dominated era starts at $z = z_{vac} \simeq 0.7$, when
$\rho_{mat} = 2 \rho_{vac}$.

It may be seen that, in vacuum-dominated universe, no
gravitational instability of the Hubble flow is possible,  and so
all deviations from the linear law cannot increase with time. This
may be seen in a simple way from the equation of motion written
in the section above.

Eq.1, after first integration, gives:
\begin{equation}
 dR/dt =
  \sqrt{(8\pi/3) \rho_{vac}}\;\; R \; \;(1 + C_1 R^{-3} + C_2 R^{-2})^{1/2},
\end{equation}
\noindent where $C_1$ is a constant proportional to the matter
mass $M= (4\pi/3) \rho R^3 = Const$, $C_2$ is a constant of
integration.

If vacuum dominates dynamically in the zero approximation, and
all other physical factors produce only relatively weak effects,
the second and third terms in the parenthesis above are small
compared to unity. These two terms give two modes of the
first-order (linear)  perturbations:
\begin{equation}
 (\delta R/R)_1 \propto R^{-3},
\end{equation}

\begin{equation}
 (\delta R/R)_2 \propto R^{-2}.
\end{equation}

Both modes are decreasing; the first one reflects the effect of
the self-gravity of matter on the matter motion, while the second
one shows the effect of the expansion on the flow of the matter
which is considered as non-gravitating, in this approximation.

There is also the third mode of linear perturbations that is
related to the possibility of arbitrary choice of the moment of
zero time; in a formal way, this mode comes from the second
integration of Eq.1 and contains the arbitrary constant $C_3$ of
this integration:
\begin{equation}
 (\delta R/R)_3 \propto C_3.
\end{equation}
This third mode is not decreasing, but "frozen".

The method of stability analysis we used here was first suggested by
Zeldovich (1965) for Lifshitz-type perturbations in the expanding Universe
with zero cosmological constant. With this method, as we see, our analysis
proves that the anti-gravity of the vacuum can change radically the
Lifshitz-Zeldovich theory of gravitational stability: only decreasing or
frozen perturbations are possible in the vacuum dominated Universe.
(For more details and also a non-linear stability analysis see
Starikova and Chernin, 2000, - in press.)

If one is interested especially in velocity perturbations, which
are deviations from the Hubble linear flow, one can find their
behaviour from the relations above. For the three modes one has
in terms of both $R$ and red-shifts:
\begin{equation}
 (\delta V)_1 \propto R^{-2} \propto (1+z)^2,
\end{equation}

\begin{equation}
 (\delta V)_2 \propto R^{-1} \propto (1+z).
\end{equation}

In both modes, the velocity perturbations are decreasing. The first one
reflects, as above, the dynamical effect of the
self-gravity of non-uniform matter on the matter motion, while the second
one shows the effect of adiabatic cooling of the flow.

In the frozen mode, the velocity perturbation is increasing:

\begin{equation}
 (\delta V)_3 \propto R \propto (1+z)^{-1}.
\end{equation}

It is easy to see, however, that the third mode is the same as the
decreasing mode of the Lifshitz-Zeldovich theory, because the both
are related to the constant of the second integration of the
equation of motion. It means that the increase of the velocity
perturbation starts from very low values in this mode, after its
big decrease during all the previous matter-dominated era. Thus,
this mode cannot introduce in reality any essential velocity
dispersion to the Hubble flow during the vacuum-dominated era.

Both the first and second modes of velocity perturbations make the flow cooler
during the cosmological expansion from  $z = z_{vac}$ to the present epoch:
\begin{equation}
\delta V/V  = (1 + z)^{n+1}, \;\; n = 1,2.
\end{equation}
>From this equation, the present-day deviations from the linear
flow are expected to be $\delta V/V \le 0.35$ for the mode with
$n=1$ and $\delta V/V \le 0.20$ for the mode with $n=2$.

It is taken here into account that the largest survived to the
present velocity perturbations were at the level $\delta V/V \le
1$, when $z = z_{vac} \simeq 0.7$; perturbations with larger
amplitudes must have formed isolated strong matter condensations
like galaxies at present, not weak deviations from the Hubble
flow.

The fact that there is no large perturbations of velocity and
density in the vacuum-dominated universe means that the process
of structure formation ceases at (or about) $z = z_{vac}$ beyond
the critical distance $R_{vac}$. In more exact words, no bound
stationary  structures can form beyond $R_{vac}$ at all, if they
did not reach nonlinear regime ($\delta V/V \simeq 1)$ at $z \ge
z_{vac}$.

We see that the vacuum is indeed an effective cooler for the
Hubble flow. However, as for the figures we get for the velocity
dispersion, they are 2-4 times larger than the observational
figure given by Sandage. And again, as in the discussion of the
section above, the discrepancy will be smaller, if one uses for
theoretical estimates a larger value of the vacuum density, --
say, $\Omega_ {vac} = 0.8-0.9$ or even  $\Omega_{vac} = 1$ (and
$\Omega_{mat} = 0.2-0.1$). May one see it as another evidence for
a larger vacuum density that comes now from the observed
quietness of the Hubble flow?

More exact numbers for the survived
velocity dispersion and probable vacuum density can be obtained only from
high-resolution large-volume cosmological simulations. Our estimates give,
most
probably, the best that can be expected in such simulations. We
mean that the simulations must take into account that
vacuum-domination regime of the flow cooling develops somewhat
slower than in our estimates, soon after $z = 0.7$, and so the
final figures for the velocity dispersion will be even somewhat larger than
given above.

\section{Cosmic intermittency}

Perhaps the higher densities of the  vacuum cannot be seen as the
only possible conclusion that one could make from the quietness of
the Hubble local flow. In general, may it indicate that, in
addition to the physics of vacuum, some new nontrivial physics
may also be involved here?

By new physics, we do not mean any revision of the fundamental
concepts, but rather ideas from other fields of physics that were
not yet exploited in cosmology. Specifically, we would like to
discuss briefly only one conjecture that may be borrowed from
modern theories of turbulence and dynamical chaos.

As is well-known, the idea of primordial cosmic chaos has deep
roots in cosmology; it has been not once supposed in modern
studies that the very early Universe was a system with complex
chaotic behaviour of the space-time and  primeval physical
fields. One example is Linde (1983)  model in which the
initial chaotic distribution of the scalar inflationary field
leads eventually to the formation of various universes with different
physics or at least different cosmological parameters. Other
interesting examples of cosmological chaos are given by
Kamenschik et al. (1998).

In a more modest version, we may consider a very early Universe in
a state of primordial chaos and assume (not necessarily within
the inflation scenario) that the structure of the initial chaos
included order as one its dynamical elements, so that order and
chaos alternated in a complex and random space-time pattern. This
conjecture is inspired by the recently developed theory of `weak
chaos' (Sagdeev {\it et al.} 1988; Zaslavsky {\it et al.} 1991),
according to which chaotic dynamics as a rule contain regularity
islands, and the weaker chaos is, the larger is the area of the
islands. This concept has proven to be fruitful in various
fields, from turbulence and plasma physics to stock market
analysis.

The phenomenon of alternation of chaos and order is called
intermittency and it is really observed in hydrodynamics. It
reveals in laboratory experiments as random bursts of turbulent
motion on the background of unstable laminar flows. Laminar and
turbulent areas are separated by irregular and moving boundaries,
and the flow velocity measured at a given point near a boundary
will show alternating laminar or turbulent motions when the point
enters or leaves the laminar region as the boundary moves.

In observations in ocean, a flow velocity measured with a device
driven with a long rope by a ship shows regular or chaotic
velocity  behaviour as the ship moves. The space pattern of the
flow may look like an archipelago of islands of laminarity in the
turbulent sea, at a given moment of time. It is this discovery of
intermittency that, in 1962, made Kolmogorov change his
1941-point of view on `Kolmogorov' turbulence. Since that
intermittency has been recognized as a robust and generic
prediction of the recent theory of chaos for a wide variety of
natural (and social) phenomena.

Suggestive insights come also from three-body problem. We studied
time series generated by three-body computer models and discovered
obvious signs of intermittency in these series (Hein\"am\"aki et
al. 1998). We observed rapid bursts of chaotic pulsations that
appear in an unpredictable manner after more or less regular
quasi-periodic behaviour of the system, and vice verse.

In the cosmological context, the weak intermittent primordial
chaos might reveal in the phase-space structure of initial
perturbations. In the phase space, "warm" areas of perturbations
with relatively higher amplitudes might alternate in chaotic
manner with "cool" areas of low amplitude perturbations, producing
together a non-stationary and evolving pattern of stochastic
behaviour. One may imagine cool spots in the warm sea or warm
spots in the cool sea of perturbations. It would be natural to
expect that the spectra of perturbations are different in these
warm and cool areas.

Intermittency can introduce specific complexity to the statistical
structure of the primordial perturbations. Some special
parameters are needed for its adequate description, which raises
new questions and demands further clarification. Everything is
complicated: the location of the cool spots in the configuration
and velocity subspaces of the phase space, their size
distribution, the shapes of the island boundaries, etc. We are
now at an early state of the art, and because of this we may
start with simplified models, in which, for instance, the
quietness of the Hubble local flow may be treated as a
manifestation of an island of quasi-regularity or a cool spot in
the generally turbulent Hubble flow.  It is on such quit spots
where one may measure locally the global Hubble constant.

Copernican principle tells us that there must be not only one and
the only  cool area in the Hubble flow, but there must be other
areas of this kind alternating with other  areas, each with its
own dispersion of the velocity field. If so, the statistical
structure of the initial cosmological perturbations must be more
complex than it is usually discussed. In particular, the initial
spectrum of the perturbations alone is not enough for the
description of this statistics in the intermittent Universe.
Perhaps cosmology will follow the evolution that the theory of
turbulence has undergone: from the universal spectrum to richness
of complexity.

\section{Vacuum and the bulk motion}

Let us turn back again to Hubble's original V vs. R plot. The
velocities and distances are given relative to the Sun, in this
plot. However, the Sun orbits around the center of the Galaxy, the
Galaxy moves in the Local Group toward the Andromeda Galaxy, the
Local Group moves toward the Local Supercluster center (the
Virgocentric flow), and, finally, there is a bulk motion of all
these masses toward the Great Attractor. The velocity of this
bulk motion is the largest one among the velocities of the other
motions in which the Sun participates: this is about 600 km/s in
the rest frame of the Cosmic Microwave Background (CMB) which the
proper cosmological reference frame. The bulk motion has a space
scale of about 100 Mpc which is comparable with the size of the
cell of uniformity. (See, for instance, Karachentsev and Makarov,
1996, 2000, for a recent discussion, new measurements and
references to original papers on the local velocity field).

It is obvious that the heliocentric reference frame used by Hubble
is not cosmological at all. Direct data on the dipole anisotropy
of the CMB indicate that the Sun moves with a velocity of about
300 km/s relative to the CMB. On the other hand, the heliocentric
velocities plotted in V vs. R diagram are the result of
superposition of various motions with velocities from 10-20 to
200-250 km/s within the whole bulk motion of all the volume. How
was Hubble able to recognize the regular cosmological flow in
this complex kinematic picture? This is another aspect of the
same Hubble enigma, now in terms of kinematics alone.

An interpretation that seems to suggest itself is that the flow
observed by Hubble is superimposed on the bulk motion, and this
superimposed flow is organized in the regular kinematic pattern
-- linear and isotropic, no matter how fast the bulk motion
against the CMB is. In other words, within the whole volume of a
hundred Mpc across, a local cosmological expansion proceeds. This
local expansion is seen as the Hubble flow and it has the same
kinematic structure and expansion rate (the Hubble constant) as
the large-scale (up to a thousand Mpc) cosmological expansion
(see Sec.3).

>From the standard cosmology viewpoint, such a local linear and
isotropic expansion within a large (100 Mpc across) volume, that
moves as a whole rapidly against the CMB, looks puzzling.
According to these standard concepts, the linear and isotropic
expansion flow on any scale can be expected only in the
cosmological reference frame that must be co-moving to the CMB.

We may, however, turn again to the idea of the cosmological
vacuum and examine if the cosmological vacuum can provide an
approach to this problem as well. Discussing the  problems of
Secs. 2-5, we referred to two properties of the cosmological
vacuum, namely its uniformity and its anti-gravity. Let us now
try the third property of the vacuum we mention in Sec.4; this is
the major mechanical property of the vacuum which makes it be
co-moving to any matter motion, or, which is the same, it makes
any matter motion to be co-moving to the vacuum.

When the vacuum dominates over the self-gravity of matter, matter
masses like galaxies and their groups move as test particles
undergoing the anti-gravity of the vacuum. The vacuum accelerates
the motions tending to form a regular kinematic pattern with
linear velocity field. The expansion rate in the flow of this
origin is the constant $H =\sqrt{ (8\pi/3) G \rho_{vac}}$, which
depends only on the value of the vacuum density.

These considerations -- in combination with the idea of cosmic
intermittency --  can be applied to a local vacuum-dominated area,
no matter how fast is the bulk motion of the area against the CMB.
They are valid as well for the large-scale matter distribution;
in both cases, the flow will be linear and the rate of expansion
will be the same on any scale. And the flow will also be stable
against any velocity or density perturbations (see Sec.5).

\section{Conclusions}

A close look at Hubble's original V vs. R plot in the light of
the present-day cosmological observations leads to a number of
seemingly simple, but non-trivial questions:

1.Does the Hubble flow really start at the very vicinity of the
Local Group, e.g. at the distance of about 2 Mpc?

2. Does the very local (2-20 Mpc) Hubble flow have indeed the same
expansion rate as the global (up to 1000 Mpc) cosmological
expansion has?

3. Does velocity dispersion in the Hubble flow is actually very
low (e.g. about 60 km/s)?

4. Is the Hubble local flow compatible in fact with the 600 km/s
bulk motion of the volume of 100 Mpc across?

We adopted above the positive answer to all these questions and
tried to explain how the idea of cosmological vacuum might provide
a physical interpretation -- at least, in principle -- of these
four properties of the  Hubble flow.

First, we demonstrated how the vacuum might induce the linear
expansion flow around us from the very small distances of 2-4 Mpc
which are much less than the size of the cosmological cell of
uniformity. This is because the vacuum starts to dominate
dynamically over dark matter at these distances.

Second, we developed stability analysis for the Hubble flow and
found that the flow could be stable, when it was controlled by the
vacuum. Because of this, the vacuum proved to be a good cooler
for the flow.

Third, we found that the flow stability could hardly explain,
however, the figure given by Sandage for the velocity
dispersion in the flow. And because of this, we suggested an
additional idea of cosmological intermittency in search for
another (or less so simple) solution of the problem.

Fourth, we sketched a possible approach to the problem of
interference of the Hubble local flow with the high-velocity bulk
motion. Both flows could be due to a complex statistical
structure of the initial intermittent perturbations and can be
compatible because the vacuum is co-moving to any matter motion
in the Universe.

Fifth, we realized that the observed quietness of the Hubble flow
and the observed figures for the minimal distance in the Hubble
law may be seen as an indication to a larger value of the vacuum
density.

Thus, the cosmological vacuum, which is uniform, anti-graviting
and universally co-moving,  dominates in the Universe since $z
\sim 1$ over all the forms of cosmic matter by density. It
makes the Universe be actually more uniform than it could be
expected from the picture of highly non-uniform matter
distribution within the observed cell of uniformity. This
uniformity reveals itself in the kinematical structure of the Hubble
matter flow which proves to be regular and quiet on the space
scales from few Mpc to 1000 Mpc. The answer we propose to the
question "Why is the Hubble flow so quiet?" is as follows: This
is most probably because the flow is mainly controlled by the
vacuum. At the same time, we feel that this may not be the final
and complete answer to the question, and one  cannot exclude that
the quietness of the Hubble flow may also be a manifestation of
some other generic physical factors or processes we do not know
now about. As an example of a new element in a possibly wider
approach to the problem, we consider the idea of cosmological
intermittency which addresses a complex statistical structure of
initial chaotic cosmological perturbations.

\section{Acknowledgements}

The work was supported in part by The Academy of Finland (project
``Cosmology in the local galaxy universe'').  We are grateful to
Chris Flynn and Igor Karachentsev for helpful comments.

{\LARGE \bf \begin{center}
References
\end{center}}

Bahcall, N.A., Lubin, L.M., Dorman, V., Where is the dark matter?,
{\em Astrophys.J.}, {\bf 447}, L81-L85, 1995.

Baryshev, Yu., Sylos Labini, F., Montuori, M., Pietronero, L.,
Teerikorpi, P.,
On the fractal structure of galaxy distribution and its implications for
cosmology, {\em Fractals} {\bf 6}, 231-243, 1998.

van den Bergh, S., The local group of galaxies.  {\em Astron.Astrophys.Rev.}
{\bf 9}, 273-318, 1999.

Ekholm, T., Lanoix, P., Teerikorpi, P., Paturel, G., Fouqu\'{e}, P.,
Investigations of the local supercluster velocity field II. A study
using Tolman-Bondi solution and galaxies with accurate distances from the
Cepheid PL-relation, {\em Astron.Astrophys.} {\bf 351}, 827-833, 1999.

Hein\"am\"aki, P., Lehto, H., Chernin A., Valtonen, M.,
Three-body dynamics: Intermittent chaos and strange attractor ,
{\em MN RAS}, {\bf 298}, 790-796, 1998

Hubble, E., A relation between distance and radial velocity among
extra-galactic nebulae, {\em Proc.Nat.Acad.Sci.}, {\bf 15}, 168-173, 1929.

Kamenschik, A.Yu., Khalatnikov, I.M., Toporensky A.V.,
Simplest cosmological models with scalar field,
{\em Int.J.Mod.Phys.} {\bf D7}, 129-137, 1998.

Karachentsev, I., Makarov, D., The Galaxy motion relative to nearby
galaxies and the local velocity field, {\em Astron.J.}, {\bf 111}, 794-803
1996.

Karachentsev, I., Makarov, D., The local velocity field of
galaxies, {\em Astrophysics}, {\bf }, 2000.

Linde, A., Chaotic inflation, {\em Phys.Lett.}, {\bf 129B}, 177-182, 1983.

Peebles, P.J.E., \underline{Principles of Cosmology}, Princeton, 1993.

Perlmuter, S. et al., Measurements of $\Omega$ and $\Lambda$ from
42 high-redshift supernovae, {\em Astrophys.J.}, {\bf 517},
565-586, 1999.

Riess, A.G. et al., Observational evidence from supernovae for an
accelerating universe and a cosmological constant, {\em Astron.J.}
 {\bf 116}, 1009-1038, 1998.

Sagdeev R.Z., Usikov D.A., Zaslavsky G.M., \underline {Nonlinear Physics:
>From the Pendulum \\
to Turbulence and Chaos}, Harwood Acad. Publ.,
New York, 1988.

Sandage, A., Bias properties of extragalactic distance indicators. VIII.
$H_0$ from distance-limited luminosity class and morphological type-specific
luminosity functions for Sb, Sbc, and Sc galaxies calibrated using cepheids,
{\em Astrophys.J.}, {\bf 527}, 479-487, 1999.

Sandage, A., The redshift-distance relation.IX. Perturbation of the
very nearby velocity field by the mass of the Local Group,
 {\em Astrophys.J.}, {\bf 307}, 1-19, 1986.

Sandage, A., Tammann, G., Hardy, E., Limits on the local deviation of
the universe from a homogeneous model, {\em Astrophys.J.}, {\bf 172},
 253-263, 1972.

Teerikorpi, P., Observational selection bias affecting the determination
of the extragalactic distance scale,  {\em Ann.Rev.Astron.Astrophys.}
{\bf 35}, 101-136, 1997.

Teerikorpi, P. et al., The radial space distribution of KLUN-galaxies
up to 200 Mpc: incompleteness or evidence for the behaviour predicted
by fractal dimension $\approx 2$ ?, {\em Astron.Astrophys.}, {\bf 334},
 395-403, 1998.

Wu, K.K.S., Lahav, O., Rees, M.J., The large-scale smoothness of
the universe,  {\em Nature}, {\bf 397}, 225-230, 1999.

Zaslavsky G.M., Sagdeev R.Z., Cherninkov A.A., Usikov D.A.,
\underline {Weak Chaos and \\
Quasiregular Patterns}, Cambridge Univ.
Press. New York, 1991.

Zehavi, I., Dekel, A., Evidence for a positive cosmological constant from
flows of galaxies and distant supernovae, {\em Nature}, {\bf 401}, 252-254,
1999.

Zeldovich Ya.B., Modern cosmology, {\em Adv.Astron.Astrophys.}, {\bf 3},
241-258, 1965.

\end{document}